\documentclass[12pt]{article}
\newcommand{\be}{\begin{equation}}
\newcommand{\ee}{\end{equation}}

\newcommand{\bi}{\begin{itemize}}
\newcommand{\ei}{\end{itemize}}
\newcommand{\bea}{\begin{eqnarray}}
\newcommand{\eea}{\end{eqnarray}}
\newcommand{\ba}{\begin{array}}
\newcommand{\ea}{\end{array}}

\usepackage[left=2.50cm, right=2.50cm, top=2.50cm, bottom=2.50cm]{geometry}
\usepackage[utf8]{inputenc}
\usepackage{cancel}
\usepackage[english]{babel}
\usepackage{physics}
\usepackage{hyperref}
\usepackage{amsthm}
\usepackage{mathrsfs}
\usepackage{mathtools}
\usepackage{graphicx}
\usepackage{changepage}
\usepackage{amssymb,amsmath}
\usepackage{verbatim}
\usepackage{empheq}
\usepackage{xcolor}
\usepackage{tikz}
\usepackage{float}
\usepackage{multirow}
\usepackage{multicol}
\usepackage{amsmath}
\usepackage{centernot}
\usepackage[hang,flushmargin]{footmisc} 
\usepackage[normalem]{ulem}
\usetikzlibrary{tikzmark}
\numberwithin{equation}{section}
\hypersetup{hidelinks}

\newlength{\bibitemsep}
\newlength{\bibparskip}
\let\oldthebibliography\thebibliography
\renewcommand\thebibliography[1]{%
  \oldthebibliography{#1}%
  \setlength{\parskip}{\bibitemsep}%
  \setlength{\itemsep}{\bibparskip}%
}
\begin{document}

\par
\bigskip
\Large
\noindent
{\bf 

Quasi-topological mass generation for 3D linearized \mbox{gravity}\\
\bigskip
\par
\rm
\normalsize

\hrule

\vspace{1cm}

\large
\noindent
{\bf Erica Bertolini$^{1,a}$},
{\bf Edoardo Lui$^{2,b}$}, 
{\bf Nicola Maggiore$^{2,3,c}$}\\

\par

\small

\noindent$^1$  School of Theoretical Physics, Dublin Institute for Advanced Studies, 10 Burlington Road, Dublin 4, Ireland.

\noindent$^2$ Dipartimento di Fisica, Universit\`a di Genova, Via Dodecaneso 33, I-16146 Genova, Italy.

\noindent$^3$ Istituto Nazionale di Fisica Nucleare - Sezione di Genova, Via Dodecaneso 33, I-16146 Genova, Italy.

\smallskip

\smallskip

\vspace{1cm}

\noindent

{\tt Abstract}\\

We present a new mass generation mechanism for linearized gravity in three spacetime dimensions, which consists of a lower-dimensional Chern-Simons-like term added to the invariant action. The propagators of the gauge fixed massive action show a massive pole and a good massless limit. Moreover, we show that, as the Topological Massive Gravity model of Deser, Jackiw and Templeton, this theory displays one propagating massive DoF, which can be traced back to the transverse part of the spatial Ricci tensor. Finally, the action of this linearized massive gravity is characterized by an algebraic structure formed by a set of Ward operators, which uniquely determine the theory.

\vspace{\fill}

\noindent{\tt Keywords:} \\
Gauge field theory, Linearized gravity, Massive gravity in 3D

\vspace{1cm}

\hrule
\noindent{\tt E-mail:
$^a$ebertolini@stp.dias.ie,
$^b$edoardo.lui@gmail.com,
$^c$nicola.maggiore@ge.infn.it.
}
\newpage

\section{Introduction and discussion of results}

The aim of this paper is to build a three-dimensional (3D) massive Linearized Gravity (LG) with the following requirements
\begin{enumerate}
\item {\bf Massless limit:} massless LG should be recovered anytime, going at vanishing mass.
\item {\bf Propagators:} the theory should have well defined propagators, with massive poles which identify the massive parameter appearing in the action as a physical mass. Dealing with a gauge field theory, propagators need a gauge fixing procedure, which depends on a gauge condition, which, in turn, follows from the gauge symmetry of the invariant action. As we will see, this latter point is a non trivial task.
\item {\bf No unphysical ghosts:} the propagators should not have tachyonic unphysical poles with negative mass. This requirement should be achieved without modifying the massless 3D LG action, which would spoil the above requirement on the existence of a good massless limit. Moreover, the absence of unphysical ghosts should not restrict the mass parameter to particular values.
\item {\bf Power counting:} the action should respect power counting, which implies the absence of the massive parameter at the denominator in the action, already forbidden by requirement 1. In 3D, gravity is a renormalizable theory \cite{Witten:2007kt}, and adding a mass should not spoil this property.
\item {\bf Degrees of Freedom (DoF):} the theory we are going to propose does not intend to be ``another'' 3D massive LG. On the contrary, it should display the same physical content of Topological Massive Gravity (TMG) discussed in \cite{Deser:1982vy,Deser:1981wh}, to which other models of 3D massive gravity make explicit reference \cite{Bergshoeff:2009hq}. In other words, we would like here to build a well defined gauge field theory of a symmetric tensor field $h_{\mu\nu}(x)$ which could rightly be called 3D massive LG.
\end{enumerate}
In order to reach our goal, as a first approach, we proceed in full analogy with the topological mass generation mechanism which characterizes Maxwell-Chern-Simons (MCS) theory, discussed in the same paper as TMG \cite{Deser:1982vy,Deser:1981wh}. To the 3D LG action $S_{\textsc{lg}}$, a lower dimensional Chern-Simons-like term $S_m$ is added, both terms being invariant under the same symmetry: the longitudinal infinitesimal diffeomorphisms \cite{Dalmazi:2020xou}
	\be\label{longsimm}
	\delta_{long} h_{\mu\nu}=\partial_\mu\partial_\nu\lambda\ ,
	\ee
which are the infinitesimal diffeomorphism transformation 
	\be
	\delta_{diff}\; h_{\mu\nu}=\partial_\mu\lambda_\nu + \partial_\nu\lambda_\mu\ ,
	\label{diffsymm}
	\ee
in the particular case
	\be
	\lambda_\mu \propto \partial_\mu\lambda\ .
	\ee
The full analogy with the MCS model relies on the fact that both actions involved, $i.e.$ the Maxwell and the Chern-Simons (CS) terms, are gauge invariant. A first, actually not that negligible, difference from MCS is that the longitudinal diffeomorphisms \eqref{longsimm} do not uniquely define the action $S_{\textsc{lg}} + S_m$, since a third invariant term exists, relevant for fractonic theories \cite{Pretko:2020cko,Pretko:2018jbi,
Bertolini:2024yur,Bertolini:2025qcy,Bertolini:2022ijb}, which one should therefore arbitrarily neglect. Besides this problem, as we will see in Section \ref{Sec3.2}, almost everything goes smoothly, in the sense that the theory respects power counting and, after a proper gauge fixing procedure, propagators are found, which display a massive pole and no tachyonic ghosts. However, propagators diverge at vanishing mass, thus violating requirement 1 of the above list. Hence, the MCS mass generating mechanism is not directly applicable for gravity. This forces us to adopt a different approach, described in Section \ref{Sec3.3}, where the infinitesimal diffeomorphisms \eqref{diffsymm} uniquely define the massless 3D LG action $S_{\textsc{lg}}$, and $S_m$ is taken as a 
massive  breaking term.
Moreover, since the gauge parameter of the transformation \eqref{diffsymm} is a vector, the gauge fixing condition and the corresponding gauge fixing term in the action must be different from the theory defined by the longitudinal diffeomorphisms \eqref{longsimm}, which are characterized by a scalar gauge parameter. This leads to different propagators, which, in this case, have a good massless limit. Quite remarkably, the resulting massive and gauge fixed LG action is characterized by an algebraic structure formed by a set of Ward operators, which uniquely determine the theory. The novelty of introducing the CS-like term $S_m$, which is not new, since it appears already in self-dual massive gravity models \cite{Aragone:1986hm, Dalmazi:2009pm, Dalmazi:2021dgp}, resides in the fact that it is the first use of that operator solely as a mass term in a quadratic Einstein-Hilbert action with a strictly
symmetric $h_{\mu \nu}(x)$ and unbroken linearised diffeomorphisms. Finally, concerning the last point 5 of our list of requirements, in Section 4 we follow the same procedure of \cite{Deser:1982vy,Deser:1981wh} based on a decomposition of the tensor field $h_{\mu\nu}(x)$ and we find that our theory has the same physical content as TMG: there is only one massive propagating DoF, which satisfies a Klein-Gordon equation, which can be traced back to the transverse part of the spatial Ricci tensor. Therefore all the required features listed at the beginning of this Section are satisfied.

The paper is organized as follows. In Section 2 we summarize the main properties of the massless 3D LG, showing in particular the absence of propagating DoF. In Section 3, after a brief description of the previous massive models, namely TMG \cite{Deser:1982vy,Deser:1981wh} and New Massive Gravity (NMG) \cite{Bergshoeff:2009hq}, we proceed to the construction of our proposal, where the full analogy with MCS theory, because of the lack of a good massless limit, makes way for the model we summarized above. The physical content (one massive propagating DoF identified in the transverse spatial Ricci tensor) is studied in Section 4.

\vspace{1cm}

{\bf Notations}

3D=2+1 spacetime dimensions .\\
Indexes:  $\mu,\nu,\rho,...=\{0,1,2\}\ \,\ i,j,k,...=\{1,2\}$ .\\
Minkowski metric: $ \eta_{\mu\nu}=\mbox{diag}(-1,1,1)\ . $ \\
Levi-Civita symbol: $\epsilon_{012}=1=-\epsilon^{012}\ $.

\section{Massless 3D LG}

The LG action
	\be\label{Slg}
	S_{\textsc{lg}}=\int d^3x\;\left(\partial_\mu h\partial^\mu h- \partial_\rho h_{\mu\nu}\partial^\rho h^{\mu\nu}-2\partial_\mu h\partial_\nu h^{\mu\nu}+2\partial_\rho h_{\mu\nu}\partial^\mu h^{\nu\rho}\right)
	\ee
is the most general local integrated functional depending on a symmetric rank-2 tensor field $h_{\mu\nu}(x)$ (whose trace is $h(x)\equiv\eta^{\mu\nu}h_{\mu\nu}(x)$), with mass dimension $[h_{\mu\nu}(x)]=1/2$, invariant under the infinitesimal diffeomorphism transformation \eqref{diffsymm}, which is a gauge transformation with a local vector parameter $\lambda_\mu(x)$. The fact that the action \eqref{Slg} (or, equivalently, the transformation \eqref{diffsymm}) defines a gauge theory manifests itself when trying to derive the propagators.  The quadratic operator $K^{\mu\nu,\rho\sigma}(x)$ defined from \eqref{Slg} by
\be
S_{\textsc{lg}}=\int d^3x\; h_{\mu\nu}K^{\mu\nu,\rho\sigma}h_{\rho\sigma}\ ,
\label{defDeltax}\ee
is not invertible, which means that, as for any other gauge theory, there is no dynamics unless a gauge condition is imposed. The customary covariant choice in LG  is the harmonic one \cite{Carroll,blau}
\be
\partial^\nu h_{\mu\nu} -\frac{1}{2}\partial_\mu h =0\ ,
\label{harmgaugecond}\ee
which, as usual in gauge field theory, does not completely fix the gauge. Indeed, requiring that all the points on the gauge orbit defined by the transformation \eqref{diffsymm} obey the gauge condition \eqref{harmgaugecond}, leaves a residual harmonic gauge condition on $\lambda_\mu(x)$
	\be
\partial^\nu h_{\mu\nu} -\frac{1}{2}\partial_\mu h =
\partial^\nu ( h_{\mu\nu} + \partial_\mu\lambda_\nu+\partial_\nu\lambda_\mu) -\frac{1}{2}\partial_\mu 
(h + 2\partial_\nu\lambda^\nu)=0
\quad\Rightarrow\quad \Box \lambda_\mu=0\ ,
\label{residual_vect}\ee
where $\Box\equiv \partial^\lambda\partial_\lambda$. This is in complete analogy with the standard abelian gauge transformation for the gauge vector field $A_\mu(x)$, for which the stability of the Lorenz gauge condition implies \cite{Itzykson}
	\be
	\partial_\mu A^\mu=\partial_\mu(A^\mu+\partial^\mu\Lambda)=0\quad\Rightarrow\quad \Box\Lambda=0\ .
	\ee
As it is well known \cite{Witten:2007kt}, despite the fact that the tensor field $h_{\mu\nu}(x)$ satisfies a wave equation, gravity in 3D has no propagating DoF. Following for instance \cite{blau}, the wave equation for $h_{\mu\nu}(x)$ is easily derived from the on-shell equation of  motion (EoM)
\be
	\frac{\delta S_{\textsc{lg}}}{\delta h^{\mu\nu}}=-2\eta_{\mu\nu}\Box h+2\eta_{\mu\nu}\partial_\alpha\partial_\beta h^{\alpha\beta}+2\partial_\mu\partial_\nu h+2\Box h_{\mu\nu}-2\partial^\alpha\left(\partial_\mu h_{\nu\alpha}+\partial_\nu h_{\mu\alpha}\right)=0\ .
\label{eomh}\ee
Taking into account the gauge condition \eqref{harmgaugecond}, the EoM \eqref{eomh} becomes
\be
-\eta_{\mu\nu}\Box h + 2 \Box h_{\mu\nu} = 0\ .
\label{eom+gc}\ee
Saturating with $\eta^{\mu\nu}$ we have
\be
\Box h=0\ ,
\label{}\ee
hence, from \eqref{eom+gc}, we get the wave equation for $h_{\mu\nu}(x)$
\be
\Box h_{\mu\nu}=0\ .
\label{wave}\ee
In 3D the wave equation \eqref{wave} does not correspond to any propagating DoF. This is easily seen following for instance  \cite{Carroll} and adopting the Transverse Traceless (TT) gauge for $h_{\mu\nu}^{TT}(x)$, which is spatial, traceless and transverse
\be
h_{0\nu}^{TT} = 
\eta^{\mu\nu}h_{\mu\nu}^{TT} = 
\partial^\mu h_{\mu\nu}^{TT} = 0\ .
\label{TTgauge}\ee
The tensor field $h_{\mu\nu}^{TT}(x)$ satisfies the wave equation $\Box h^{TT}_{\mu\nu}(x)=0$, which is solved by 
\be
h_{\mu\nu}^{TT} = C_{\mu\nu}e^{ik_\lambda x^\lambda}\ .
\label{Cmunu}\ee
On the TT-gauge \eqref{TTgauge}, it is immediate to see that, requiring \eqref{TTgauge}, the constant coefficients $C_{\mu\nu}$ must vanish
\be
C_{\mu\nu}=0\ .
\label{C=0}\ee

\section{Massive 3D LG}\label{Sec3}

\subsection{Topological massive gravity and new massive gravity}\label{Sec3.1}

The main massive gravity models in 3D whose linear approximation provides for a mass to the graviton are two \cite{Hinterbichler:2011tt}, which we are going to briefly summarize in what follows: TMG \cite{Deser:1982vy,Deser:1981wh} and NMG \cite{Bergshoeff:2009hq}.

\paragraph{TMG} Following the same approach of MCS theory \cite{Deser:1982vy,Deser:1981wh}, where the lower dimensional CS term gives a mass to the vector field in 3D, in that same paper a CS-like term is coupled to the Einstein-Hilbert (EH) action as follows
	\be
	S_{\textsc{tmg}}=\tfrac{M_P}{2}\int d^3x\left[-\sqrt{-g}\;\;R-\tfrac{1}{2\mu_{\textsc t}}\epsilon^{\lambda\mu\nu}\Gamma^\alpha_{\lambda\beta}\left(\partial_\mu\Gamma^\beta_{\alpha\nu}+\tfrac{2}{3}\Gamma^\beta_{\mu\gamma}\Gamma^\gamma_{\nu\alpha}\right)\right]\ ,
\label{Stmg}	\ee
where $M_P$ is the Planck mass and $\Gamma^\mu_{\rho\sigma}(x)$ is the Christoffel connection. As the ordinary CS action, the CS-like term in \eqref{Stmg} breaks parity and, 
even though it is not a covariant tensor, the full action preserves diffeomorphisms invariance up to boundary terms \cite{Hinterbichler:2011tt}. Differently from the CS term in the MCS theory, the CS-like term in \eqref{Stmg} strongly depends on the spacetime metric $g_{\mu\nu}(x)$, through the Christoffel symbols  $\Gamma^\mu_{\rho\sigma}(x)$ and the presence of contracted indices. Nevertheless it provides for a mass generation mechanism similar to the topological one of the MCS theory \cite{Deser:1982vy,Deser:1981wh}. In fact from the EoM
	\be\label{eomTMG}
	\mathcal O^{\mu\alpha}(\mu_{\textsc t})\Box h^{TT}_{\alpha\beta}=0\ ,
	\ee
where
\be
 \mathcal O^{\mu\alpha}(\mu_{\textsc t})\equiv\eta^{\mu\alpha}+\tfrac{1}{\mu_{\textsc t}}\epsilon^{\mu\lambda\alpha}\partial_\lambda\ ,
	\ee
one can get, by multiplying by $\mathcal O(-\mu_{\textsc t})$ \cite{Hinterbichler:2011tt}, the following Klein-Gordon equation
	\be
	(\Box-\mu_{\textsc t}^2)\Box h_{\mu\nu}^{TT}=0\ .
	\ee
In \cite{Deser:1982vy,Deser:1981wh} it is shown that the linearized TMG action reduces to
	\be\label{TMG-KG}
	S_{\textsc{tmg}}^{lin}\sim\int d^3x\, \varphi(\Box-\mu_{\textsc t}^2)\varphi\ ,
	\ee
where the scalar field $\varphi(x)$ is the transverse component of the metric perturbation $h_{\mu\nu}(x)$. This shows that TMG is characterized by one single propagating spin-2 DoF with mass $\mu_{\textsc t}$ \cite{Deser:1982vy,Deser:1981wh}.

\paragraph{NMG} Presented in \cite{Bergshoeff:2009hq}, this theory describes a parity-preserving massive theory for the graviton, characterized by two propagating DoF as the massless 4D case, but differently from TMG, which {describes one massive propagating spin-2 DoF only, represented by the scalar function $\varphi(x)$ in \eqref{TMG-KG}}. The action of NMG is
	\be
	S_{\textsc{nmg}}=\tfrac{M_P}{2}\int d^3x\sqrt{-g}\;\;\left[-R+\tfrac{1}{\mu^2_{\textsc n}}\left(R_{\mu\nu}R^{\mu\nu}-\tfrac{3}{8}R^2\right)\right]\ ,
\label{nmg}	\ee
where  $\mu_{\textsc n}$ is a massive coefficient at the denominator of the higher order term added to the EH action. Notice that, as in TMG, the EH term has the {``wrong'' sign (as pointed out by the authors of NMG themselves)}: in both cases this has been done in order to avoid ghost-like physical states \cite{Bergshoeff:2009hq}.
The linearized EoM are given by
	\be\label{eomNMG}
	(\Box-\mu^2_{\textsc n})G_{\mu\nu}^{lin}=0\quad;\quad R^{lin}=0\ ,
	\ee
where $G_{\mu\nu}^{lin}(x)$ and $R^{lin}(x)$ are the linearized Einstein tensor and the linearized Ricci scalar curvature, respectively. In \cite{Bergshoeff:2009hq} it is shown that the above EoM imply that the physical propagating massive DoF are two, corresponding to the transverse traceless components of the metric perturbation, with helicity $\pm2$ and mass $\mu_{\textsc n}$. \\

TMG and NMG give mass to the 3D graviton in two different ways. The mass term in TMG is the gravitational extension of the topological mass generation applied to ordinary gauge theory in 3D and, due to the presence of the Levi-Civita antisymmetric tensor $\epsilon^{\mu\nu\rho}$, is peculiar to 3D spacetime. On the other hand, NMG reminds the Fierz-Pauli mass term of LG \cite{Hinterbichler:2011tt}
\be
\int d^3x\sqrt{-g}\;
\tfrac{1}{\mu^2_{\textsc n}}\left(R_{\mu\nu}R^{\mu\nu}-\tfrac{3}{8}R^2\right)
\longleftrightarrow
\int d^3x\;
m^2\left(h_{\mu\nu}h^{\mu\nu}-h^2\right)\ ,
\label{nmg-fp}\ee
which is not constrained to a particular spacetime dimension. 
Differently from the MCS theory, the mass term of both TMG and NMG have higher mass dimensions (three and four, respectively) than the massless EH term (two), which implies the presence of a dimensional parameter at the denominator in both cases: $\tfrac{1}{\mu_{\textsc t}}$ in TMG and $\tfrac{1}{\mu^2_{\textsc n}}$ in NMG. This  fact has two consequences. The  first is the absence of a good massless limit in both cases: the massless EH action is recovered (albeit with the {``wrong''} sign) for $\mu\to\infty$. Concerning that, we remark that the limit for large graviton mass does not correspond to the physical situation which, on the contrary, rather leans towards a tiny mass for the graviton ($\mu\rightarrow 0$), where both TMG and NMG diverge, as in the Fierz-Pauli theory, which lacks well defined massless propagators \cite{Blasi:2017pkk,Blasi:2015lrg,Gambuti:2020onb,Gambuti:2021meo}. Related to this, the presence of a mass parameter at the denominator allows for infinite invariant terms in the action, which therefore is not uniquely constrained by a symmetry, as, on the other hand, General Relativity, which indeed must be considered as an effective theory. Moreover and finally, in both theories it is necessary to set the coefficients of the model, including the EH term to the {``wrong''} sign, in order to avoid ghost-like physical states.

\subsection{Quasitopological mass generation}\label{Sec3.2}

We propose here an alternative way of giving a mass to 3D LG, using MCS theory \cite{Deser:1982vy,Deser:1981wh} as a toy model for the mass generation mechanism, without introducing power counting violating higher derivatives terms in the action, and keeping its covariant feature as well. Following the same approach described in \cite{Deser:1982vy,Deser:1981wh}, we add to the LG action \eqref{Slg} the lower dimensional term 
	\be
	S_{m}=\int d^3x\,\epsilon^{\mu\nu\rho} h_{\mu}^{\;\lambda}\partial_\nu h_{\rho\lambda}\label{massterm}\ ,
	\ee
which, when considered alone, encodes a higher rank CS theory of fractons, whose Hall-like behaviour has been described in \cite{Bertolini:2024yur}. 
{Notice that the presence in $S_m$ \eqref{massterm} of the Levi-Civita tensor $\epsilon^{\mu\nu\rho}$ induces the breaking of parity, as in TMG}.
The analogy  with  the ordinary abelian CS term is evident
	\be
	S_{CS}=\int d^3x\,\epsilon^{\mu\nu\rho} A_\mu\partial_\nu A_\rho \label{CS}\ .
	\ee
The full action is thus
	\be
	S=S_{\textsc{lg}}+mS_{m}\ ,\label{vectaction}
	\ee
where $m$ is a constant massive {real} parameter $[m]=1$. Notice that for vanishing $m$, the LG action \eqref{Slg} is trivially recovered. The action $S$ \eqref{vectaction} is invariant under the longitudinal infinitesimal diffeomorphisms \eqref{longsimm}.
As $S_{\textsc{lg}}$ \eqref{defDeltax}, the free quadratic operator in the action $S$ \eqref{vectaction} is not invertible and the propagator cannot be computed until the scalar gauge fixing condition
\be
\kappa_0\partial^\mu\partial^\nu h_{\mu\nu} + \kappa_1\Box h=0\ ,
\label{}\ee
which corresponds to the scalar transformation \eqref{longsimm}, is imposed.
This is done in Appendix \ref{Slg+Sminv+Sgf},
where we expanded the momentum space propagator 
\be
\langle\Tilde{h}_{\alpha\beta}(p)\Tilde{h}_{\rho\sigma}(-p)\rangle = \sum_{i=0}^6 c_i(p) A^{(i)}_{\alpha\beta,\rho\sigma}(p)\; , 
\label{hhpropcongf}\ee
on the basis $\{A^{(i)}_{\alpha\beta,\rho\sigma}(p)\}$ \cite{Bertolini:2024yur}
\bea
A^{(0)}_{\alpha\beta,\rho\sigma}  &=& \frac{1}{2}(\eta_{\alpha\rho}\eta_{\beta\sigma}+\eta_{\alpha\sigma}\eta_{\beta\rho}) 
\label{A0}\\
A^{(1)}_{\alpha\beta,\rho\sigma} &=& 
\frac{1}{p^2}(\eta_{\alpha\rho}p_\beta p_\sigma +  
\eta_{\alpha\sigma}p_\beta p_\rho +
\eta_{\beta\rho}p_\alpha p_\sigma +
\eta_{\beta\sigma}p_\alpha p_\rho)
\label{A1}\\
A^{(2)}_{\alpha\beta,\rho\sigma} &=& 
\frac{1}{p^2}(\eta_{\alpha\beta} p_\rho p_\sigma +
\eta_{\rho\sigma} p_\alpha p_\beta)
\label{A2} \\
A^{(3)}_{\alpha\beta,\rho\sigma} &=& 
\eta_{\alpha\beta}\eta_{\rho\sigma}
\label{A3}\\
A^{(4)}_{\alpha\beta,\rho\sigma} &=& 
\frac{p_\alpha p_\beta p_\rho p_\sigma}{p^4}
\label{A4}\\
A^{(5)}_{\alpha\beta,\rho\sigma} &=& 
ip^\lambda (
\epsilon_{\alpha\lambda\rho} \eta_{\sigma\beta} +
\epsilon_{\beta\lambda\rho} \eta_{\sigma\alpha} +
\epsilon_{\alpha\lambda\sigma} \eta_{\rho\beta} +
\epsilon_{\beta\lambda\sigma} \eta_{\rho\alpha}
)
\label{A5}\\
A^{(6)}_{\alpha\beta,\rho\sigma} &=&
 \frac{ip^\lambda}{p^2} (
\epsilon_{\alpha\lambda\rho} p_\sigma p_\beta +
\epsilon_{\alpha\lambda\sigma} p_\rho p_\beta +
\epsilon_{\beta\lambda\rho} p_\sigma p_\alpha +
\epsilon_{\beta\lambda\sigma} p_\rho p_\alpha
)\; .
\label{A6}
\eea
The coefficients $c_i$ are found to be
\bea 
c_0 &=& -\frac{1}{p^2 + m^2}\label{firstscalsol}\\ 
c_1 &=& \frac{1}{2(p^2+m^2)} \\ 
c_2 &=& -\frac{1}{2}\biggl[\frac{\kappa_0+3\kappa_1}{p^2(\kappa_0+\kappa_1)} +\frac{1}{p^2+m^2}\biggr]\\
c_3 &=&\frac{1}{2}\left(\frac{1}{p^2} +\frac{1}{p^2+m^2}\right)\\ 
c_4 &=& \frac{1}{2}\biggl[\frac{(\kappa_0+3\kappa_1)^2}{p^2(\kappa_0+\kappa_1)^2} -\frac{1}{p^2+m^2}\biggr]\\ 
c_5 &=& \frac{m}{4p^2(p^2+m^2)}\\ 
c_6 &=& \frac{1}{4p^2}\left(\frac{4}{m}-\frac{m}{p^2+m^2} \right) \label{c6div}\ .  
\eea
We notice that the massless limit of the tensor field propagator in momentum space \eqref{hhpropcongf}
diverges (because of the coefficient $c_6$ \eqref{c6div}), despite the fact that, as we said, $S_{\textsc{lg}}$ is recovered from $S$ for $m\rightarrow 0$. The reason for this non intuitive result rests on the gauge structure of the theory, which strongly depends on $m$. The actions $S_{\textsc{lg}}$ \eqref{Slg} and $S$ \eqref{vectaction} are invariant under different transformations: the pure LG action $S_{\textsc{lg}}$ is invariant under the infinitesimal diffeomorphisms \eqref{diffsymm}, which depends on the $vector$ gauge parameter $\lambda_\mu(x)$, while the massive action $S$ \eqref{vectaction} is invariant under the subset of longitudinal diffeomorphisms \eqref{longsimm}, whose gauge parameter is the $scalar$ $\lambda(x)$. Therefore the massless limit of the theory changes the symmetry from \eqref{longsimm} to \eqref{diffsymm}, and, consequently, the gauge condition must be modified from scalar to vector and the gauge fixing term must change accordingly. This does not happen in MCS theory, since both Maxwell and CS actions are invariant under the same gauge transformation. Here the situation is interestingly different, and the massless limit of the theory appears to be a more subtle issue. This drives us to the conclusion that, in order to have a massive LG with smooth massless limit, the mechanism strictly analogous to the topological mass generation \cite{Deser:1982vy,Deser:1981wh} cannot be straightforwardly applied.

\subsection{BRS formulation of 3D massive linearized gravity}\label{Sec3.3}

The standard way to gauge fix a gauge field theory is given by the BRS procedure \cite{Becchi:1975nq,Stora:1976kd}, according to which the local gauge parameter $\lambda_\mu(x)$ is promoted to a ghost quantum field $c_\mu(x)$, and an antighost $\bar c_\mu(x)$ and a Nakanishi-Lautrup multiplier $b_\mu(x)$ \cite{Nakanishi:1966zz,Lautrup:1967zz} are introduced as well. The BRS field transformations are defined as follows
	\bea
	 sh_{\mu\nu}&=&\partial_\mu c_\nu+\partial_\nu c_\mu\label{s0h}\\
	 sc_\mu&=&0\\
	s\bar c_\mu&=&b_\mu\\
sb_\mu&=&0\ .
	\eea
The BRS operator $s$ is nilpotent
\be
s^2=0\ ,
\label{nilpotency}\ee
and is a symmetry of the LG action $S_{\textsc{lg}}$ \eqref{Slg}
\be
sS_{\textsc{lg}}=0\ ,
\label{s0SLG}\ee
which is obvious, since the BRS variation \eqref{s0h} of the tensor field $h_{\mu\nu}(x)$ coincides with the diffeomorphism transformation \eqref{diffsymm}. As usual, the gauge fixing term is a BRS cocycle
\be
S_{gf} = s\int d^3x\; \bar c^\mu \left(  \partial^\nu h_{\mu\nu} -\frac{1}{2}\partial_\mu h\right)
=
\int d^3x\; \left[ b^\mu \left(\partial^\nu h_{\mu\nu} -\frac{1}{2}\partial_\mu h\right)
-\bar c^\mu\Box c_\mu\right]   \label{brsSgf}\ ,
\ee
where the fact that the BRS operator $s$ and the antighost field $\bar c^\mu(x)$ are anticommuting Grassmann variables has been taken into account. The harmonic gauge condition \eqref{harmgaugecond} is enforced by the EoM of the Lagrange multiplier $b^\mu(x)$ and, due to the symmetry \eqref{s0SLG} and the nilpotency \eqref{nilpotency}, the gauge fixed action is BRS invariant
\be
s(S_{\textsc{lg}}+S_{gf})= 0\ .
\label{}\ee
The CS-like mass term $S_m$ \eqref{massterm} is not invariant under diffeomorphisms, hence it breaks the BRS symmetry $s$. However, by modifying the transformation of the multiplier $b_\mu(x)$, we get the new BRS operator $s_m$
	\bea
	 s_m h_{\mu\nu}&=&\partial_\mu c_\nu+\partial_\nu c_\mu\\
	 s_m c_\mu&=&0\\
	s_m \bar c_\mu&=&b_\mu\\
s_m b_\mu&=&2m\;\epsilon_{\mu\nu\rho}\partial^\nu c^\rho\ ,
	\eea
which, remarkably, is an $exact$ symmetry of the massive gauge fixed action
\be
s_m S_{tot}=0\ ,
\label{sStot}\ee
with
	\be
	S_{tot}\equiv S_{\textsc{lg}}+S_{gf}+S_m\ ,\label{Stot}
	\ee
as it can be easily verified. The presence of the mass term $S_m$ \eqref{massterm} spoils the nilpotency of the BRS operator $s_m$, opening however the doors to an interesting algebraic structure of the massive theory
\be
s_m^2=\delta\ \ ;\ \
[s_m,\delta]=\delta^2=0\ ,
\label{algebra}\ee
where the operator $\delta$ acts on the antighost $\bar c_\mu(x)$ only
\be
\delta\bar c_\mu=2m\;\epsilon_{\mu\nu\rho}\partial^\nu c^\rho\ ,
\label{deltabarc}\ee
and is a symmetry of the massive theory
\be
\delta S_{tot}=0\ .
\label{deltasimm}\ee
Both the BRS and $\delta$ symmetries of the total action \eqref{Stot}  can be written in a functional way by means of the corresponding Ward operators as follows
\bea
\hat s_m S_{tot} &=&
 \int d^3x\;
\left[\left(\partial_\mu c_\nu+\partial_\nu c_\mu\right)
\frac{\delta}{\delta h_{\mu\nu}}
+b_\mu\frac{\delta}{\delta\bar c_\mu}
+2m\epsilon_{\mu\nu\rho}\partial^\nu c^\rho \frac{\delta}{\delta b_\mu}
\right]S_{tot}=0 \label{wardsm}\\
\hat\delta S_{tot} &=&
\int d^3x\;
\epsilon_{\mu\nu\rho}\partial^\nu c^\rho\frac{\delta }{\delta\bar c_\mu}\;S_{tot} =0\ .
\label{warddelta}
\eea
A comment is in order concerning the ghost sector in the gauge fixing term $S_{gf}$ \eqref{brsSgf}. Despite the fact that $c_\mu(x)$ and $\bar c_\mu(x)$ are decoupled from the other fields and might be integrated out from the path integral, as in standard abelian gauge theory where the ghost fields do not play any role, in our case their presence is needed in order to have the symmetry of the massive theory \eqref{sStot}. 
The remarkable outcome is that $S_{m}$ \eqref{massterm} serves indeed as a mass term, since a massive pole of the tensor field propagator \eqref{hhpropcongf} is found which, differently from the case considered in Section \ref{Sec3.2}, displays a good massless limit. In fact, 
in Appendix  \ref{Slg+Smnoinv+Sgf} we show that the coefficients appearing in the propagator \eqref{hhpropcongf} in this case are
\bea 
c_0& =&-\frac{1}{p^2 + m^2}\label{firstvectex}\\
c_1& =& \frac{1}{2(p^2 + m^2)}\\ 
c_2& =& \frac{m^2}{2p^2(p^2 + m^2)}\\
c_3& =&\frac{1}{2}\biggl(\frac{1}{p^2}+\frac{1}{p^2+m^2}\biggr)\\
c_4& =& \frac{m^2}{2p^2(p^2 + m^2)}\\ 
c_5& =& \frac{m}{4p^2(p^2+m^2)}\\
c_6& =&  -\frac{m}{4p^2(p^2+m^2)}\ . 
 \label{lastvectex}\eea
 Remarkably, adopting the gauge choice of \cite{Deser:1982vy,Deser:1981wh}, these propagators coincide with those of TMG, despite the fact that the massive term is different. In particular, the pole structure of the propagator \eqref{hhpropcongf} coincides with that in \cite{Deser:1982vy,Deser:1981wh}: we see indeed the presence of two poles, at $p^2=0$ and $p^2=-m^2$. {In order that the massive pole is physical, we need that $m^2>0$. Given the tensor field propagator \eqref{hhpropcongf}, it is useful to write it in the Barnes-River basis in position space \cite{Nakasone:2009vt,Accioly:2012yz}
\begin{align}
	P^{(2)}_{\mu\nu,\alpha\beta}&\equiv\tfrac{1}{2}\left(\theta_{\mu\alpha}\theta_{\nu\beta}+\theta_{\nu\alpha}\theta_{\mu\beta}-\theta_{\mu\nu}\theta_{\alpha\beta}\right)\label{P2}\\
	P^{(1)}_{\mu\nu,\alpha\beta}&\equiv\tfrac{1}{2}\left(\theta_{\mu\alpha}\omega_{\nu\beta}+\theta_{\nu\alpha}\omega_{\mu\beta}+\theta_{\mu\beta}\omega_{\nu\alpha}+\theta_{\nu\beta}\omega_{\mu\alpha}\right)\\
	P^{(0,s)}_{\mu\nu,\alpha\beta}&\equiv\tfrac{1}{2}\theta_{\mu\nu}\theta_{\alpha\beta}\\
	P^{(0,w)}_{\mu\nu,\alpha\beta}&\equiv\omega_{\mu\nu}\omega_{\alpha\beta}\\
	P^{(0,sw)}_{\mu\nu,\alpha\beta}+P^{(0,ws)}_{\mu\nu,\alpha\beta}&\equiv\tfrac{1}{\sqrt2}\left(\theta_{\mu\nu}\omega_{\alpha\beta}+\theta_{\alpha\beta}\omega_{\mu\nu}\right)\label{P0}\\
	\mathcal E^{(\theta)}_{\mu\nu,\alpha\beta}&\equiv  (\epsilon_{\alpha\lambda\rho} \theta_{\sigma\beta} +\epsilon_{\beta\lambda\rho}\theta_{\sigma\alpha} +\epsilon_{\alpha\lambda\sigma} \theta_{\rho\beta} +\epsilon_{\beta\lambda\sigma} \theta_{\rho\alpha})\partial^\lambda\label{Et}\\
	\mathcal E^{(w)}_{\mu\nu,\alpha\beta}&\equiv  (\epsilon_{\alpha\lambda\rho} \omega_{\sigma\beta} +\epsilon_{\beta\lambda\rho}\omega_{\sigma\alpha} +\epsilon_{\alpha\lambda\sigma} \omega_{\rho\beta} +\epsilon_{\beta\lambda\sigma} \omega_{\rho\alpha})\partial^\lambda\ ,\label{Ew}
	\end{align}
where the $P(x)$ operators \eqref{P2}-\eqref{P0} are spin projectors for a symmetric rank-2 tensor field. For instance $P^{(2)}(x)$ is associated to spin-2, and is thus particularly relevant in gravity theories. The operators $\mathcal E^{(\theta)}(x)$ \eqref{Et} and $\mathcal E^{(w)}(x)$ \eqref{Ew} complete the basis due to the existence of nontrivial Levi-Civita terms in the 3D case, and finally
	\begin{equation}
	\theta_{\mu\nu}\equiv \eta_{\mu\nu}-\omega_{\mu\nu}\quad;\quad\omega_{\mu\nu}\equiv\frac{1}{\Box}\partial_\mu\partial_\nu\ 
	\end{equation}
are transverse and longitudinal projectors respectively.
Following for instance the analysis of \cite{Nakasone:2009bn,Nakasone:2009vt}, the Barnes-River basis allows to isolate spin contributions, identifying some properties of the theory such as unitarity, positive energy and absence of ghosts. The basis \eqref{P2}-\eqref{Ew} can be therefore used in our case instead of the $\{A^{(i)}(p)\}$  \eqref{A0}-\eqref{A6}, so that the tensor field propagator \eqref{hhpropcongf} becomes, in position space
\begin{equation}
		\Delta_{\mu\nu,\alpha\beta}=
		\left\{
		\frac{1}{2}\frac{1}{\Box-m^2}\left(P^{(2)}-\frac{m}{2\Box}\mathcal E^{(\theta)}\right)
		-\frac{1}{\Box}\left[\sqrt2\left(P^{(0,ws)}+P^{(0,sw)}\right)+P^{(0,s)}+2P^{(0,w)}\right]
		\right\}_{\mu\nu,\alpha\beta}\ .
	\end{equation}
This shows that the massive pole is associated to the spin-2 projector $P^{(2)}(x)$ \eqref{P2}, and describes a unitary mode of positive mass $m^2$ and positive norm. No massive ghosts appear. One can compare this result with the one obtained, for instance, in \cite{Nakasone:2009vt}, where a massive gravity theory is studied by considering both TMG \eqref{Stmg} and NMG \eqref{nmg} contributions and a Fierz-Pauli mass term \eqref{nmg-fp}. 
In that case it becomes evident that  the remark about the  ``wrong" sign in front of the LG contribution made in \cite{Deser:1982vy,Deser:1981wh,Bergshoeff:2009hq} comes from requiring that the energy density is positive, which is not granted due to the presence in the action of higher derivatives terms, absent in our case. Another remark concerns the sign of the CS-like term, which is related, as in TMG, to the helicity $\pm 2$ \cite{Deser:1982vy,Deser:1981wh,Bergshoeff:2009hq}}.  In the next Section we will interpret and identify the propagating massive DoF. Concerning the massless DoF, it does not propagate, since the massless poles disappear when the propagator is contracted into conserved sources, as explicitly shown in \cite{Deser:1982vy,Deser:1981wh}. As a final comment,  we stress the ``natural'' absence of the vDVZ discontinuity \cite{vanDam:1970vg,Zakharov:1970cc} in our model. In fact, it  is associated with the absence of a smooth massless limit in LG with a Fierz-Pauli mass term. The root of the problem lies in the pathological dual use of the Fierz-Pauli term, which simultaneously acts as both a mass term and a gauge-fixing term. However, when a Fierz-Pauli mass term is added to LG in four dimensions already properly gauge fixed, the vDVZ discontinuity is absent. This has been explicitly shown in \cite{Blasi:2017pkk, Blasi:2015lrg, Gambuti:2020onb, Gambuti:2021meo}. In our case, not only do we give mass to the graviton without using a Fierz-Pauli term, but we do so via a CS-like action constructed from a symmetric tensor. Moreover, our action is dynamically well defined due to the presence of a proper gauge-fixing procedure. Therefore, the issue of the vDVZ discontinuity, which is specifically linked to the use of a Fierz-Pauli mass term and the resulting lack of a regular massless limit, simply does not arise in our framework.

\section{Gravitational waves and massive propagating DoF}

To further confirm that the theory we are proposing here is a candidate for massive LG in 3D, in this Section we show that the action $S$ \eqref{vectaction}, with $S_m$ \eqref{massterm} as a mass term, possesses one propagating scalar DoF, in accordance with what was shown in \cite{Deser:1982vy,Deser:1981wh}. To reach this goal, we adopt the approach followed in \cite{Deser:1982vy,Deser:1981wh}, which basically consists in decomposing the tensor field $h_{\mu\nu}(x)$  into irreducible representations of the rotation group, which are further decomposed into their transverse, longitudinal and, for the symmetric spatial tensor, solenoidal components. Explicitly this means
\begin{align}
h_{00}&=\phi\\
h_{0i}&=u_i+\partial_i\omega\\
h_{ij} &= h_{ij}^T + h_{ij}^L + h_{ij}^S \nonumber \\
&=
[(\eta_{ij}-\hat\partial_i\hat\partial_j)\varphi] +
[\hat\partial_i\hat\partial_j\chi] +
[\partial_i\xi_j+\partial_j\xi_i]\ , \label{TLS}
\end{align}
where
\be
\hat\partial_i \equiv \frac{\partial_i}{\sqrt{\nabla^2}}\ ,
\label{}\ee
and the spatial vector fields $u_i(x)$ and $\xi_i(x)$  are transverse
\bea
\partial^i u_i &=&0 
\label{transvu}\\
\partial^i\xi_i &=&0\ ,
\label{transvxi}
\eea
so that $h_{ij}^T(x)$, $h_{ij}^L(x)$ and $h_{ij}^S(x)$ represent the transverse, longitudinal and solenoidal parts of $h_{ij}(x)$, respectively~:
\bea
\partial^i h_{ij}^T &=& 0 \\
\epsilon^{0ij}\partial_i\partial^k h_{jk}^L &=& 0  \\
\partial^i\partial^j h_{ij}^S &=& 0\ .
\eea
The action $S$ \eqref{vectaction} can be written in terms of the fields appearing in the decomposition of $h_{\mu\nu}(x)$ as
\begin{align}
S &= \int d^3x\; \left\{
2\phi\nabla^2\varphi 
-2 \varphi\partial_0^2\chi
-4\varphi\partial_0\nabla^2\omega
-2u^i\nabla^2 u_i
+2\xi^i\partial_0^2\nabla^2\xi_i
+4u^i\partial_0\nabla^2\xi_i \right.  \label{axdecomposta}\\
&\qquad
\left.
+m\epsilon^{0ij}\left[
-2(\phi+\varphi)\partial_iu_j
-2\xi_i\partial_j\nabla^2\omega
+u_i\partial_0u_j
-2\omega\partial_0\partial_iu_j
+2(\varphi-\chi)\partial_0\partial_i\xi_j
+\xi_i\partial_0\nabla^2\xi_j
\right]
\right\}\;.\nonumber
\end{align}
Due to the transversality conditions \eqref{transvu} and \eqref{transvxi}, the vectors $u_i(x)$ and $\xi_i(x)$ can be expressed in terms of scalar fields
\be
u_i=\epsilon_{0ij}\partial^ju\ \ ;\ \ 
\xi_i=\epsilon_{0ij}\partial^j\xi\ ,
\ee
so that the action \eqref{axdecomposta} can be written entirely in terms of scalar fields
\bea
S &=& 2\int d^3x\; \left[
\nabla^2\phi(\varphi-mu)
+\chi (\partial_0^2\varphi-m\partial_0\nabla^2\xi)
+\nabla^2\omega(2\partial_0\varphi-m\nabla^2\xi-m\partial_0u)
 \right. \nonumber \\
&&\left.
\quad +(\nabla^2u)^2
+(\partial_0\nabla^2\xi-2\nabla^2u)   \partial_0\nabla^2\xi
-m\varphi(\nabla^2u
-\partial_0\nabla^2\xi)
\right]\ . \label{scalaraction}
\eea
From \eqref{scalaraction} we see that $\phi(x)$, $\omega(x)$ and $\chi(x)$ are Lagrange multipliers for the constraints
\bea
\nabla^2(\varphi-mu) &=& 0\label{poisson1} \\
\nabla^2(2\partial_0\varphi-m\nabla^2\xi-m\partial_0u) &=& 0 \label{poisson2}\\
\partial_0^2\varphi-m\partial_0\nabla^2\xi &=&0\ . \label{nonpoisson3}
\eea
In particular, \eqref{poisson1} and \eqref{poisson2} are Poisson equations which, assuming that the fields vanish at infinity, are solved by
\bea
\varphi &=& mu \label{solpoisson1}\\
\nabla^2\xi &=& -\partial_0(u -\frac{2}{m}\varphi) =\partial_0u \ ,\label{solpoisson2}
\eea
where in \eqref{solpoisson2} we used \eqref{solpoisson1}. Using \eqref{solpoisson1} and \eqref{solpoisson2}, the action S \eqref{scalaraction} reduces to
\be
S=\int d^3x\; \left[ u (\Box-m^2)\Box u\right]\ ,
\label{kgaction}\ee
which shows that $\Box u(x)$ is a scalar massive DoF which propagates through the massive Klein-Gordon equation
\be
(\Box-m^2)\Box u =0\ .
\label{KG}\ee
This result agrees with \cite{Deser:1982vy,Deser:1981wh}, with the nice difference that the 3D massive LG presented here has a good massless limit, is power-counting renormalizable and, above all, is invariant under the BRS symmetry \eqref{sStot}. In order to identify the tensorial quantity that propagates as a massive wave, we observe that the three spatial components of the Ricci tensor can be identified in the same way as the metric perturbation \eqref{TLS}, in terms of two scalar fields $\mathcal R^T(x)$ and $\mathcal R^L(x)$, and a transverse vector field $\mathcal R^S_i(x)$
	\be
	R_{ij}\equiv R_{ij}^T+R_{ij}^L+R_{ij}^S=(\eta_{ij}-\hat\partial_i\hat\partial_j)\mathcal R^T+\hat\partial_i\hat\partial_j\mathcal R^L+\partial_i\mathcal R_j^S+\partial_j\mathcal R_i^S\ .\label{Rdecomp}
	\ee
On the other hand, the linearized approximation of the Ricci tensor, in the harmonic gauge \eqref{harmgaugecond}, is
	\be
	R_{\mu\nu}\propto\Box h_{\mu\nu}\ .\label{R=Boxh}
	\ee
Hence, by applying the projector $\eta^{ij}-\hat\partial^i\hat\partial^j$ to the $ij$-components of \eqref{R=Boxh} and using the solution \eqref{solpoisson1}, we can identify the scalar propagating DoF $\Box u(x)$ in the Klein-Gordon equation \eqref{KG} with the scalar component $\mathcal R^T(x)$ of $R_{ij}(x)$ in \eqref{Rdecomp}
	\be
	\mathcal R^T\,\longleftrightarrow\,\Box u\ .
	\label{identification}\ee
It is thus the transverse component of the Ricci tensor{, which is a spin-2 field,} that propagates as a massive wave, $i.e.$
	\be
	(\Box-m^2)R^T_{ij} =0\ .\label{massiveR}
	\ee
As a confirmation of the identification \eqref{identification}, we remark that in \cite{Hinterbichler:2011tt} the scalar propagating DoF of \cite{Deser:1982vy,Deser:1981wh} is identified with \mbox{$\Box h_{\mu\nu}^{TT}(x)\sim R_{\mu\nu}(x)$}, where $h_{\mu\nu}^{TT}(x)$ is the tensor field in the transverse traceless gauge described in Section 2. {This is the same result of TMG : the transverse component of the spin-2 spatial Ricci tensor, which is described by a scalar field,  propagates as a massive wave. In other words: we are dealing with a single massive mode of helicity $\pm 2$, as it should. From \eqref{massiveR} it is clear that the propagating DoF has spin 2, which is described by a single mode \eqref{KG}. }

\section*{Acknowledgments}

We thank Camillo Imbimbo for enlightening remarks concerning the counting of the degrees of freedom for generic tensor field theories.

\appendix

\section{Propagators}

In order to compute the propagators of a gauge field theory, a gauge condition must be imposed. The tensorial character of the gauge condition  depends on the type of transformation. For instance, to scalar gauge parameters correspond scalar gauge conditions, like the ordinary gauge transformation of a vector gauge field and the longitudinal infinitesimal diffeomorphisms on a symmetric rank-2 tensor field~: \\
\textbf{gauge}
\begin{align}
\delta_{gauge}A_\mu = \partial_\mu\Lambda\quad &\rightarrow\quad \partial_\mu A^\mu=0\ ;
\label{gaugeapp}
\intertext{\bf longitudinal diff}
\delta_{long}h_{\mu\nu} = \partial_\mu\partial_\nu\lambda \quad&\rightarrow \quad
\kappa_0\partial^\mu\partial^\nu h_{\mu\nu} + 
\kappa_1\Box  h =0\ .
\label{longdiffapp}
\intertext{On the other hand, to vector gauge parameters correspond vector gauge conditions, and an example is given by the infinitesimal diffeomorphism transformation}\intertext{\bf diff}
\delta_{diff}h_{\mu\nu} = \partial_\mu\lambda_\nu + \partial_\nu\lambda_\mu
\quad &\rightarrow \quad
k_0\partial^\nu h_{\mu\nu} + k_1\partial_\mu h =0\ .
\label{diffapp}\end{align}
The gauge conditions \eqref{longdiffapp} and \eqref{diffapp} depend on the constant parameters 
$(\kappa_0,\kappa_1)$  and $(k_0,k_1)$, and are implemented by adding gauge fixing terms to the invariant action which, in $D$ spacetime dimensions, are
\bea
S_{gf}^{(gauge)} &=& 
-\frac{1}{2\xi}\int d^Dx\; (\partial_\mu A^\mu)^2
\\
S_{gf}^{(long)} &=& 
-\frac{1}{2\xi}\int d^Dx\; 
(\kappa_0\partial^\mu\partial^\nu h_{\mu\nu} + \kappa_1\Box  h)^2
\\
S_{gf}^{(diff)} &=& 
-\frac{1}{2\xi}\int d^Dx\; 
(k_0\partial^\nu h_{\mu\nu} + k_1\partial_\mu h)^2\ ,
\eea
where the $\xi$'s are the gauge parameters which tune, for instance, the Landau gauge $(\xi=0)$ or the Feynman gauge $(\xi=1)$. It is often convenient to linearize the gauge fixing term by means of a Lagrange multiplier, a.k.a. Nakanishi-Lautrup field \cite{Nakanishi:1966zz,Lautrup:1967zz}. In that case, which is the one we adopt here, the above gauge fixing terms read
\bea
S_{gf}^{(gauge)} &=& 
\int d^Dx\; b\left(\partial_\mu A^\mu + \frac{\xi}{2}b\right)
\\
S_{gf}^{(long)} &=& 
\int d^Dx\; 
b\left(\kappa_0\partial^\mu\partial^\nu h_{\mu\nu} + \kappa_1\Box  h + \frac{\xi}{2}b\right)
\label{Sgflong}\\
S_{gf}^{(diff)} &=& 
\int d^Dx\; 
b^\mu\left(k_0\partial^\nu h_{\mu\nu} + k_1\partial_\mu h+\frac{\xi}{2}b_\mu\right)\ . \label{Sgfdiff}
\eea
In the following Subsections we compute the propagators in the Landau gauge $\xi=0$ for the two cases we discuss in this paper.  

\subsection{Scalar gauge fixing for longitudinal infinitesimal diffeomorphisms}\label{Slg+Sminv+Sgf}

In this case the total (Landau) gauge fixed action is
\be
S=\left.S_{\textsc{lg}}+mS_m+S^{(long)}_{gf}\right|_{\xi=0}\ ,
\label{totactfract}\ee
where both $S_{\textsc{lg}}$ \eqref{Slg} and $S_m$ \eqref{massterm} are invariant under the longitudinal diffeomorphisms \eqref{longsimm}, which need the scalar gauge fixing condition \eqref{longdiffapp} and the corresponding gauge fixing action term \eqref{Sgflong}. In momentum space\footnote{The Fourier transform is defined
$h_{\mu\nu}(x)=\int d^3p\, e^{ip_\lambda x^\lambda}\Tilde{h}_{\mu\nu}(p)$.} the action $S$ \eqref{totactfract} reads
	\begin{align}
	S &= \int d^3p\; \left\{\tilde{h}_{\mu\nu}(p)\left[-p^2\eta^{\mu\alpha}\eta^{\nu\beta}
	+2\eta^{\mu\alpha} p^\nu p^\beta 
	-2\eta^{\mu\nu} p^\alpha p^ \beta+p^2\eta^{\mu\nu} \eta^{\alpha\beta}
	-mi p_\lambda 
	\epsilon^{\mu\lambda\alpha} \eta^{\beta\nu}\right]\tilde{h}_{\alpha\beta}(-p)\right.\nonumber\\
	&\quad\quad\quad\left. +\; \tilde{b}(p)\left(-\kappa_0p^\alpha p^\beta- \kappa_1 p^2 \eta^{\alpha\beta}\right)\tilde{h}_{\alpha\beta}(-p)\right\}\nonumber\\
	&\equiv\int d^3p\; \tilde{\phi}_M(p)\tilde{K}^{MA}(p)\tilde{\phi}_A(-p)\ ,\label{momS}
	\end{align}
with
\be
\tilde{\phi}_M(p) \equiv  \left(\tilde{h}_{\mu\nu}(p), \; \tilde{b}(p)\right)\label{defphivect1}
\ee
and
\be
{\tilde{K}}^{MA}(p)  \equiv \begin{bmatrix}
{\tilde{K}}^{\mu\nu,\alpha\beta}(p) & \mathit{\tilde{K}}^{*\,\mu\nu}(p)\\ 
  \mathit{\tilde{K}}^{\alpha\beta}(p) & 
  0
\end{bmatrix} \label{defKvect1}\ .
\ee
Due to the fact that $h_{\mu\nu}(x)$ is a symmetric tensor field, the following relations on the quadratic operators appearing in \eqref{defKvect1} hold
\begin{align} 
\tilde{K}^{\mu\nu,\alpha\beta}(p) &= \tilde{K}^{\nu\mu,\alpha\beta}(p) = \tilde{K}^{\mu\nu,\beta\alpha}(p)=\tilde{K}^{\alpha\beta,\mu\nu}(-p)=\tilde{K}^{*\,\mu\nu,\alpha\beta}(p) \\
\tilde{K}^{\alpha\beta}(p) &= \tilde{K}^{\beta\alpha}(p) \ ,
\label{simmK}\end{align}
 and the tensor $\tilde{K}^{\mu\nu,\alpha\beta}(p)$ can be expanded on the basis $\{A^{(i)}(p)\}$ \eqref{A0}-\eqref{A6}
as
\begin{align}
\tilde{K}^{\mu\nu,\alpha\beta}& =
\left[p^2\left(
-A^{(0)}+\frac{1}{2}A^{(1)}{\tiny-A^{(2)}+A^{(3)}}\right){\tiny -\frac{m}{4}A^{(5)}}\right]^{\mu\nu,\alpha\beta}\\
\tilde{K}^{\alpha\beta}& =-\frac{1}{2}(\kappa_0p^\alpha p^\beta+ \kappa_1 p^2 \eta^{\alpha\beta})\  .
\label{Kexpansion}\end{align}
The matrix of the propagators in momentum space is
\be
\tilde{\Delta}_{AP}(p) \equiv \begin{bmatrix}
 \tilde{\Delta}_{\alpha\beta,\rho\sigma}(p) & \tilde{\Delta}^*_{\alpha\beta}(p)\\ 
  \tilde{\Delta}_{\rho\sigma}(p) & \tilde{\Delta}(p)
\end{bmatrix} \ ,
\label{defDelta}\ee

with
\begin{align} 
 \tilde{\Delta}_{\alpha\beta,\rho\sigma}(p)&\equiv \langle\Tilde{h}_{\alpha\beta}(p)\Tilde{h}_{\rho\sigma}(-p)\rangle  \\ 
 \tilde{\Delta}_{\alpha\beta}(p)&\equiv \langle\Tilde{h}_{\alpha\beta}(p) \tilde{b}(-p)\rangle\\
 \tilde{\Delta}(p)&\equiv \langle\tilde{b}(p) \tilde{b}(-p)\rangle\; ,
\end{align}
such that
\be
{\tilde{K}}^{MA}(p)\tilde{\Delta}_{AP}(p)=\begin{bmatrix}
 {\mathcal{I}}^{\mu\nu}_{\rho\sigma} & 0\\ 
  0 & 1
\end{bmatrix} \ ,
\label{defprop}\ee
where ${\mathcal{I}}^{\mu\nu}_{\rho\sigma}$ is the identity
\be
{\cal I}^{\mu\nu}_{\rho\sigma} = \frac{1}{2}(\delta^\mu_\rho\delta^\nu_\sigma + \delta^\mu_\sigma\delta^\nu_\rho)\; .
\label{identity}\ee
On the basis $\{A^{(i)}(p)\}$ \eqref{A0}-\eqref{A6} the propagator $\tilde{\Delta}_{\alpha\beta,\rho\sigma}(p)$ can be expanded as
\be
\tilde{\Delta}_{\alpha\beta,\rho\sigma}(p) = \sum_{i=0}^6 c_i(p) A^{(i)}_{\alpha\beta,\rho\sigma}(p)\; , 
\label{Delta1base}\ee
and
\be
\tilde{\Delta}_{\alpha\beta}(p) = a_1(p)\eta_{\alpha\beta} + a_2(p)\frac{p_\alpha p_\beta}{p^2}\label{Delta2base} \ ,
\ee
where $c_i(p)$ and $a_1(p)$, $a_2(p)$ are real functions. Our aim is to evaluate $c_i(p)$, $a_1(p)$, $a_2(p)$ from \eqref{defprop}, which explicitly reads:
\bea\label{propagatorrelations}
&&\tilde{K}^{\mu\nu,\alpha\beta}\tilde{\Delta}_{\alpha\beta,\rho\sigma} + \tilde{K}^{*\; \mu\nu}\tilde{\Delta}_{\rho\sigma} = {\cal I}^{\mu\nu}_{\rho\sigma}\\
&&\tilde{K}^{\mu\nu,\alpha\beta}\tilde{\Delta}^*_{\alpha\beta} + \tilde{K}^{\alpha\beta}\tilde{\Delta} = 0\label{propagatorrelation2}\\
&&\tilde{K}^{\alpha\beta}\tilde{\Delta}_{\alpha\beta,\rho\sigma} = 0 \label{propagatorrelation3}\\
&&\tilde{K}^{\alpha\beta}\tilde{\Delta}^*_{\alpha\beta} = 1 \; .
\label{propagatorrelation4}\eea

After long but straightforward calculations, the solution of the propagator equation \eqref{defprop} is
\bea 
c_0 &=& -\frac{1}{p^2 + m^2}\label{firstscalsolApp}\\ 
c_1 &=& \frac{1}{2(p^2+m^2)} \\ 
c_2 &=& -\frac{1}{2}\biggl[\frac{\kappa_0+3\kappa_1}{p^2(\kappa_0+\kappa_1)} +\frac{1}{p^2+m^2}\biggr]\\
c_3 &=&\frac{1}{2}\left(\frac{1}{p^2} +\frac{1}{p^2+m^2}\right)\\ 
c_4 &=& \frac{1}{2}\biggl[\frac{(\kappa_0+3\kappa_1)^2}{p^2(\kappa_0+\kappa_1)^2} -\frac{1}{p^2+m^2}\biggr]\\ 
c_5 &=& \frac{m}{4p^2(p^2+m^2)}\\ 
c_6 &=& \frac{1}{4p^2}\left(\frac{4}{m}-\frac{m}{p^2+m^2} \right)\label{c6}\\
a_1 &=& 0 \\
a_2&=&-\frac{2}{p^2(\kappa_0 + \kappa_1)}\label{lastscalsolApp}\; ,
\eea
which represents the expressions of the coefficients in the propagator expansions \eqref{Delta1base} and \eqref{Delta2base}, together with $\tilde{\Delta}(p)=0$. We remark the presence of the massive pole 
\be
p^2=-m^2\; .
\ee
 However, we observe as well that there is no good massless limit, notably because of the $c_6$-coefficient \eqref{c6} in the expansion \eqref{Delta1base} of the tensor field propagator $\tilde{\Delta}_{\alpha\beta,\rho\sigma}(p)$. This result is expected, since  for $m\to 0$ the theory reduces to pure LG, which is invariant under the more general infinitesimal diffeomorphism transformation \eqref{diffsymm}, characterized by the vector gauge parameter $\lambda_\mu(x)$, which requires a vector gauge-fixing condition instead of the scalar one \eqref{Sgflong} we used. 

\subsection{Vector gauge fixing for infinitesimal diffeomorphisms}\label{Slg+Smnoinv+Sgf}

We deal here with our proposal for 3D massive LG: $S_m$ \eqref{massterm} breaks the invariance of $S_{\textsc{lg}}$ \eqref{Slg} under the infinitesimal diffeomorphisms \eqref{diffsymm}, 
but the gauge fixed massive action \eqref{Stot} satisfies the BRS symmetry \eqref{sStot}.
The gauge fixing condition is the harmonic one \eqref{harmgaugecond}, which falls into the class \eqref{diffapp} with $k_0=1$ and $k_1=-\frac{1}{2}$. The corresponding gauge fixing term, in the Landau gauge,  is
\be
S_{gf}
=\int d^3x \; b^\mu\biggl(\partial^\nu h_{\mu\nu} -\frac{1}{2}\partial_\mu h\biggr)\; .
\label{S_gflg}\ee

In momentum space, the total (massive and gauge fixed) action $S_{tot}$ \eqref{Stot} reads
\begin{align}
S_{tot}&=S_{\textsc{lg}}+ mS_{m}+S_{gf}\nonumber\\
&= \int d^3p \left[ \tilde{h}_{\mu\nu}(p)\left(-p^2\eta^{\mu\alpha}\eta^{\nu\beta}+2\eta^{\mu\alpha} p^\nu p^\beta -2\eta^{\mu\nu} p^\alpha p^ \beta +p^2\eta^{\mu\nu} \eta^{\alpha\beta}+im p_\lambda 
\epsilon^{\mu\lambda\alpha} \eta^{\beta\nu}\right)\tilde{h}_{\alpha\beta}(-p)\right.\nonumber\\
&\quad\quad\left.- \frac{i}{2}\tilde{b}_\mu(p)\left(p^\alpha \eta^{\mu\beta}+ p^\beta \eta^{\mu\alpha}- p^\mu \eta^{\alpha\beta}\right)\tilde{h}_{\alpha\beta}(-p)\right]
\nonumber\\
&\equiv\int d^3p\; \tilde{\phi}_M(p)\tilde{K}^{MA}(p)\tilde{\phi}_A(-p)\ ,\label{momSg1}
\end{align}
now with
\be
\tilde{\phi}_M(p) \equiv  \biggl(\tilde{h}_{\mu\nu}(p), \; \tilde{b}_\mu(p)\biggr)\label{defphivect}
\ee
and
\be
{\tilde{K}}^{MA}(p)  \equiv \begin{bmatrix}
 \tilde{K}^{\mu\nu,\alpha\beta}(p) & \tilde{K}^{*\; \mu\nu,\alpha}(p)\\ 
  \tilde{K}^{\mu,\alpha\beta}(p) & 0
\end{bmatrix} \label{defKvect}\ ,
\ee

where the tensor operators in \eqref{defKvect} are expanded on the basis 
$\{A^{(i)}(p)\}$ \eqref{A0}-\eqref{A6} as
\bea
\tilde{K}^{\mu\nu,\alpha\beta}& = &
\left[p^2\left(
-A^{(0)}+\frac{1}{2}A^{(1)}-A^{(2)}+A^{(3)}\right)-\frac{m}{4}A^{(5)}\right]^{\mu\nu,\alpha\beta}\\
\tilde{K}^{\mu,\alpha\beta}&=&-\frac{i}{4}\left(p^\alpha \eta^{\mu\beta}+ p^\beta \eta^{\mu\alpha}- p^\mu \eta^{\alpha\beta}\right)
\eea
with
\be
\tilde{K}^{\mu,\alpha\beta}(p) = \tilde{K}^{\mu,\beta\alpha}(p)\ .
\ee
The matrix of the propagators in momentum space is
\be
\tilde{\Delta}_{AP}(p) \equiv \begin{bmatrix}
 \tilde{\Delta}_{\alpha\beta,\rho\sigma}(p) & \tilde{\Delta}^*_{\rho,\alpha\beta}(p)\\ 
  \tilde{\Delta}_{\alpha,\rho\sigma}(p) & \tilde{\Delta}_{\alpha\rho}(p)
\end{bmatrix} \ ,
\label{defDeltavect}\ee
with
\begin{align} 
 \tilde{\Delta}_{\alpha\beta,\rho\sigma}(p)&=\langle\Tilde{h}_{\alpha\beta}(p)\Tilde{h}_{\rho\sigma}(-p)\rangle  \label{propfieldvec}\\ 
 \tilde{\Delta}_{\alpha\beta,\rho}(p)&=\langle\Tilde{h}_{\alpha\beta}(p) \tilde{b}_{\rho}(-p)\rangle\\
 \tilde{\Delta}_{\alpha\rho}(p)&= \langle\tilde{b}_\alpha(p) \tilde{b}_\rho(-p)\rangle \ ,
\end{align}
and such that
\be
{\tilde{K}}^{MA}(p)\tilde{\Delta}_{AP}(p)=\begin{bmatrix}
 {\mathcal{I}}^{\mu\nu}_{\rho\sigma} & 0\\ 
  0 & \delta^{\mu}_{\rho}
\end{bmatrix} \ ,
\label{defpropvect}\ee
where ${\mathcal{I}}^{\mu\nu}_{\rho\sigma}$ is the tensor identity \eqref{identity}. The propagator $\tilde{\Delta}_{\alpha\beta,\rho\sigma}(p)$ \eqref{propfieldvec} can be expanded on the basis $\{A^{(i)}(p)\}$ \eqref{A0}-\eqref{A6}
\be
\tilde{\Delta}_{\alpha\beta,\rho\sigma}(p) = \sum_{i=0}^6 c_i(p) A^{(i)}_{\alpha\beta,\rho\sigma}(p)\; , 
\label{Delta1basevec}\ee
and
\bea
\tilde{\Delta}_{\mu,\alpha\beta}(p) &=& ia_1(p)(\eta_{\mu\beta}p_\alpha +\eta_{\mu\alpha}p_\beta) + ia_2(p)\eta_{\alpha\beta}p_\mu + ia_3(p)p_\beta p_\alpha p_\mu + \nonumber\\
&&+ \;a_4(p)p^\lambda(\epsilon_{\mu\lambda\alpha}p_\beta+\epsilon_{\mu\lambda\beta}p_\alpha) \\
\tilde{\Delta}_{\alpha\rho}(p) &=& a_5(p)\eta_{\alpha\rho} + a_6(p)p_\alpha p_\rho + ia_7(p)p^\lambda\epsilon_{\alpha\lambda\rho}\ , 
\eea
where $c_i(p)$ and $a_i(p)$ are real functions determined by \eqref{defpropvect}, which reads
\bea
&&\tilde{K}^{\mu\nu,\alpha\beta}\tilde{\Delta}_{\alpha\beta,\rho\sigma} + \tilde{K}^{*\; \mu\nu,\alpha}\tilde{\Delta}_{\alpha,\rho\sigma}= {\cal I}^{\mu\nu}_{\rho\sigma}\label{propagatorrelationsvect}\\
&&\tilde{K}^{\mu\nu,\alpha\beta}\tilde{\Delta}^*_{\alpha\beta,\rho} + \tilde{K}^{*\,\mu\nu,\alpha}\tilde{\Delta}_{\alpha\rho} =\; 0\label{proprelvect2}\\
&&\tilde{K}^{\mu,\alpha\beta}\tilde{\Delta}_{\alpha\beta,\rho\sigma}+\tilde{K}^{\mu\alpha}\tilde{\Delta}_{\alpha,\rho\sigma}= 0\label{proprelvect3}\\
&&\tilde{K}^{\mu,\alpha\beta}\tilde{\Delta}^*_{\alpha\beta,\rho} +\tilde{K}^{\mu\alpha}\tilde{\Delta}_{\alpha\rho}= \delta^{\mu}_{\rho}\label{propagatorrelation4vect}\; .
\eea
After lengthy calculations one can show that the non vanishing coefficients appearing in 
$\tilde{\Delta}_{AP}(p)$ which solves
\eqref{defpropvect} are
\bea 
c_0& =&-\frac{1}{p^2 + m^2}\label{firstvectex1}\\
c_1& =& \frac{1}{2(p^2 + m^2)}\\ 
c_2& =& \frac{m^2}{2p^2(p^2 + m^2)}\\
c_3& =&\frac{1}{2}\biggl(\frac{1}{p^2}+\frac{1}{p^2+m^2}\biggr)\\
c_4& =& \frac{m^2}{2p^2(p^2 + m^2)}\\
c_5& =& \frac{m}{4p^2(p^2+m^2)}\\
c_6& =&  -\frac{m}{4p^2(p^2+m^2)}\\
a_1& =& -\frac{2}{p^2}\\ 
a_7& =& 4\frac{m}{p^2}\; . 
 \label{lastvectex}\eea
The presence of the massive pole 
\be
 p^2=-m^2\; \ ,
 \label{polevect}\ee
and the good massless limit of the non vanishing coefficients \eqref{firstvectex1}-\eqref{lastvectex}
\be 
c_0 =-\frac{1}{p^2}\;,\quad c_1 = \frac{1}{2p^2}\;,\quad c_3 =\frac{1}{p^2}\;,\quad a_1 = -\frac{2}{p^2}\; \ ,
\ee
render the theory described in Section \ref{Sec3.3} a good candidate for massive LG in 3D. 



\begin{thebibliography}{15}

\bibitem{Witten:2007kt}
E.~Witten,
``Three-Dimensional Gravity Revisited,''
[arXiv:0706.3359 [hep-th]].

\bibitem{Deser:1982vy}
S.~Deser, R.~Jackiw and S.~Templeton,
``Three-Dimensional Massive Gauge Theories,''
Phys. Rev. Lett. \textbf{48}, 975-978 (1982)
doi:10.1103/PhysRevLett.48.975.

\bibitem{Deser:1981wh}
S.~Deser, R.~Jackiw and S.~Templeton,
``Topologically Massive Gauge Theories,''
Annals Phys. \textbf{140}, 372-411 (1982)
[erratum: Annals Phys. \textbf{185}, 406 (1988)]
doi:10.1016/0003-4916(82)90164-6.

\bibitem{Bergshoeff:2009hq}
E.~A.~Bergshoeff, O.~Hohm and P.~K.~Townsend,
``Massive Gravity in Three Dimensions,''
Phys. Rev. Lett. \textbf{102}, 201301 (2009)
doi:10.1103/PhysRevLett.102.201301.

\bibitem{Dalmazi:2020xou}
D.~Dalmazi and R.~R.~L.~d.~Santos,
``The dimensional reduction of linearized spin-2 theories invariant under transverse diffeomorphisms,''
Eur. Phys. J. C \textbf{81}, no.6, 547 (2021)
doi:10.1140/epjc/s10052-021-09297-0.

\bibitem{Pretko:2020cko}
M.~Pretko, X.~Chen and Y.~You,
``Fracton Phases of Matter,''
Int. J. Mod. Phys. A \textbf{35}, no.06, 2030003 (2020)
doi:10.1142/S0217751X20300033.

\bibitem{Pretko:2018jbi}
M.~Pretko,
``The Fracton Gauge Principle,''
Phys. Rev. B \textbf{98}, no.11, 115134 (2018)
doi:10.1103/PhysRevB.98.115134.

\bibitem{Bertolini:2024yur}
E.~Bertolini, A.~Blasi, N.~Maggiore and D.~S.~Shaikh,
``Hall-like behaviour of higher rank Chern-Simons theory of fractons,''
JHEP \textbf{10}, 232 (2024)
doi:10.1007/JHEP10(2024)232.

\bibitem{Bertolini:2025qcy}
E.~Bertolini, A.~Blasi and N.~Maggiore,
``Quasi-topological fractons: a 3D dipolar gauge theory,''
Eur. Phys. J. C \textbf{85}, no.1, 68 (2025)
doi:10.1140/epjc/s10052-025-13821-x.

\bibitem{Bertolini:2022ijb}
E.~Bertolini and N.~Maggiore,
``Maxwell theory of fractons,''
Phys. Rev. D \textbf{106}, no.12, 125008 (2022)
doi:10.1103/PhysRevD.106.125008.


\bibitem{Aragone:1986hm}
C.~Aragone and A.~Khoudeir,
``Selfdual massive gravity,''
Phys. Lett. B \textbf{173}, 141-144 (1986)
doi:10.1016/0370-2693(86)90234-0

\bibitem{Dalmazi:2009pm}
D.~Dalmazi and E.~L.~Mendonca,
``A New spin-2 self-dual model in D=2+1,''
JHEP \textbf{09}, 011 (2009)
doi:10.1088/1126-6708/2009/09/011
[arXiv:0907.5009 [hep-th]].

\bibitem{Dalmazi:2021dgp}
D.~Dalmazi and A.~L.~R.~d.~Santos,
``Higher spin analogs of linearized topologically massive gravity and linearized new massive gravity,''
Phys. Rev. D \textbf{104}, no.8, 085023 (2021)
doi:10.1103/PhysRevD.104.085023
[arXiv:2107.08879 [hep-th]].


\bibitem{Carroll}
Carroll, S. M., (2019). ``Spacetime and Geometry: An Introduction to General Relativity.'' Cambridge University Press. 
ISBN: 9781108488396
doi.org/10.1017/9781108770385.


\bibitem{blau}
M.~Blau,
``Lecture notes on general relativity'',
\url{http://blau.itp.unibe.ch/GRLecturenotes.html}.

\bibitem{Itzykson}
C.~Itzykson, J.~B.~Zuber,
``Quantum Field Theory'',
Dover Publications (2012),
ISBN: 9780486445687.

\bibitem{Hinterbichler:2011tt}
K.~Hinterbichler,
``Theoretical Aspects of Massive Gravity,''
Rev. Mod. Phys. \textbf{84}, 671-710 (2012)
doi:10.1103/RevModPhys.84.671.

\bibitem{Blasi:2017pkk}
A.~Blasi and N.~Maggiore,
``Massive gravity and Fierz-Pauli theory,''
Eur. Phys. J. C \textbf{77}, no.9, 614 (2017)
doi:10.1140/epjc/s10052-017-5205-y.

\bibitem{Blasi:2015lrg}
A.~Blasi and N.~Maggiore,
``Massive deformations of rank-2 symmetric tensor theory (a.k.a. BRS characterization of Fierz\textendash{}Pauli massive gravity),''
Class. Quant. Grav. \textbf{34}, no.1, 015005 (2017)
doi:10.1088/1361-6382/34/1/015005.

\bibitem{Gambuti:2020onb}
G.~Gambuti and N.~Maggiore,
``A note on harmonic gauge(s) in massive gravity,''
Phys. Lett. B \textbf{807}, 135530 (2020)
doi:10.1016/j.physletb.2020.135530.

\bibitem{Gambuti:2021meo}
G.~Gambuti and N.~Maggiore,
``Fierz\textendash{}Pauli theory reloaded: from a theory of a symmetric tensor field to linearized massive gravity,''
Eur. Phys. J. C \textbf{81}, no.2, 171 (2021)
doi:10.1140/epjc/s10052-021-08962-8.

\bibitem{Becchi:1975nq}
C.~Becchi, A.~Rouet and R.~Stora,
``Renormalization of Gauge Theories,''
Annals Phys. \textbf{98}, 287-321 (1976)
doi:10.1016/0003-4916(76)90156-1.

\bibitem{Stora:1976kd}
R.~Stora,
``Continuum Gauge Theories,''
Conf. Proc. C \textbf{7607121}, 201 (1976)
CPT-76-P.854.

\bibitem{Nakanishi:1966zz}
N.~Nakanishi,
``Covariant Quantization of the Electromagnetic Field in the Landau Gauge,''
Prog. Theor. Phys. \textbf{35} (1966), 1111-1116
doi:10.1143/PTP.35.1111.

\bibitem{Lautrup:1967zz} 
  B.~Lautrup,
 ``Canonical Quantum Electrodynamics In Covariant Gauges,''
  Kong.\ Dan.\ Vid.\ Sel.\ Mat.\ Fys.\ Med.\  {\bf 35}, no. 11 (1967).
  

\bibitem{Nakasone:2009vt}
M.~Nakasone and I.~Oda,
``Massive Gravity with Mass Term in Three Dimensions,''
Phys. Rev. D \textbf{79} (2009), 104012
doi:10.1103/PhysRevD.79.104012.

\bibitem{Accioly:2012yz}
A.~Accioly, J.~Helayel-Neto, B.~Pereira-Dias and C.~Hernaski,
``A new class of spin projection operators for 3D models,''
Phys. Rev. D \textbf{86} (2012), 105046
doi:10.1103/PhysRevD.86.105046.


\bibitem{Nakasone:2009bn}
M.~Nakasone and I.~Oda,
``On Unitarity of Massive Gravity in Three Dimensions,''
Prog. Theor. Phys. \textbf{121} (2009), 1389-1397
doi:10.1143/PTP.121.1389.


\bibitem{vanDam:1970vg}
H.~van Dam and M.~J.~G.~Veltman,
``Massive and massless Yang-Mills and gravitational fields,''
Nucl. Phys. B \textbf{22}, 397-411 (1970)
doi:10.1016/0550-3213(70)90416-5

\bibitem{Zakharov:1970cc}
V.~I.~Zakharov,
``Linearized gravitation theory and the graviton mass,''
JETP Lett. \textbf{12}, 312 (1970)
\normalcolor


\end{thebibliography}
\end{document}